\documentstyle[epsfig]{aipproc}

\begin{document}
\begin{flushright}
\hfill {\rm UR-1621}\\
\hfill {\rm ER/40685/959}\\
\hfill {\rm January 2001}\\
\end{flushright}
\vspace*{2\baselineskip}

\title{Progress Towards a Generator for BFKL Physics}

\author{Lynne H.~Orr$^*$ and W.J~Stirling$^{\dagger}$}
\address{$^*$Department of Physics and Astronomy\\
University of Rochester\\
Rochester, NY 14627-0171\\
$^{\dagger}$Institute for Particle Physics Phenomenology\\ 
University of Durham\\
Durham DH1 3LE, UK}

\maketitle

\begin{abstract}

In certain regions of phase space in jet production, 
large logarithms can arise which are resummed by the BFKL equation.  
Linear colliders can potentially be  excellent places to study BFKL 
effects in jet production.  We discuss an approach to BFKL
calculations which incorporates kinematic effects explicitly and
can be implemented in an event generator.

\end{abstract}

\section*{Introduction}

In certain kinematic regimes for some QCD processes, each power of the 
strong coupling constant $\alpha_s$ is accompanied by a large logarithm
and fixed order perturbation theory fails.  To calculate anything meaningful,
the large logs must be resummed.  The BFKL equation \cite{bfkl} resums
these large logs --- which are due to multiple real and virtual gluon 
emissions --- for the regime where the gluons have comparable transverse 
momenta but are strongly ordered in rapidity.  

This BFKL physics comes into play at high energy linear $e^+e^-$ colliders 
in virtual photon scattering.  The electron and positron emit 
virtual photons, which then scatter, producing multiple jets.  
The leading order  QCD process is $\gamma^*\gamma^*\to q\bar{q}q\bar{q}$
via gluon exchange, and the corresponding BFKL process has a gluon 
ladder attached to the $t$-channel gluon.
BFKL
applies when the invariant mass $W$ of the hadronic system is
large, and $$s>>Q^2>>\Lambda_{QCD}^2,$$ where $s$ is the square of the 
center of mass energy and $Q^2$ is the virtuality of the photons.  
Physically, this corresponds to $e^+e^- \to e^+e^- +\ {\rm hadrons}$,
where the final electron and positron are scattered at low angles
(forward scattering).

\section*{Improved BFKL Approach}

The BFKL equation can be solved analytically in very special circumstances.
The analytic solutions involve summing over arbitrary numbers of gluons and 
integrating over arbitrarily large gluon transverse energies.  Only
leading order kinematics are included and there is  no kinematic
cost to emit gluons.   It is difficult to reach appropriately
asymptotic regions in experiments, so it is no surprise that 
BFKL predictions often overshoot data.  

A solution to this problem has been obtained  by solving
the BFKL equation iteratively, making the gluon sum explicit so that 
kinematic cuts can be applied directly and the solution can be implemented
in a Monte Carlo program \cite{os}; see also \cite{schmidt}.  
The BFKL equation contains separate integrals over  real and virtual 
emitted gluons.  We can reorganize the equation by combining the 
`unresolved' real emissions --- those with transverse momenta
below some minimum value (in practice chosen to be small
compared to the momentum threshold for measured
jets) --- with the virtual emissions.  
We  perform
the integration over virtual and unresolved real
emissions  analytically.  The integral containing the 
resolvable real emissions is left explicit.  The resulting subprocess
cross section is
\begin{equation}
d\hat\sigma=d\hat\sigma_0\times\sum_{n\ge 0} f_{n}
\end{equation}
where $f_{n}$ is the iterated solution for $n$ real gluons emitted and
contains an overall form factor due
to virtual and unresolved emissions.

\begin{figure}
\epsfxsize200pt
\vspace*{-1cm}
\psfig{figure=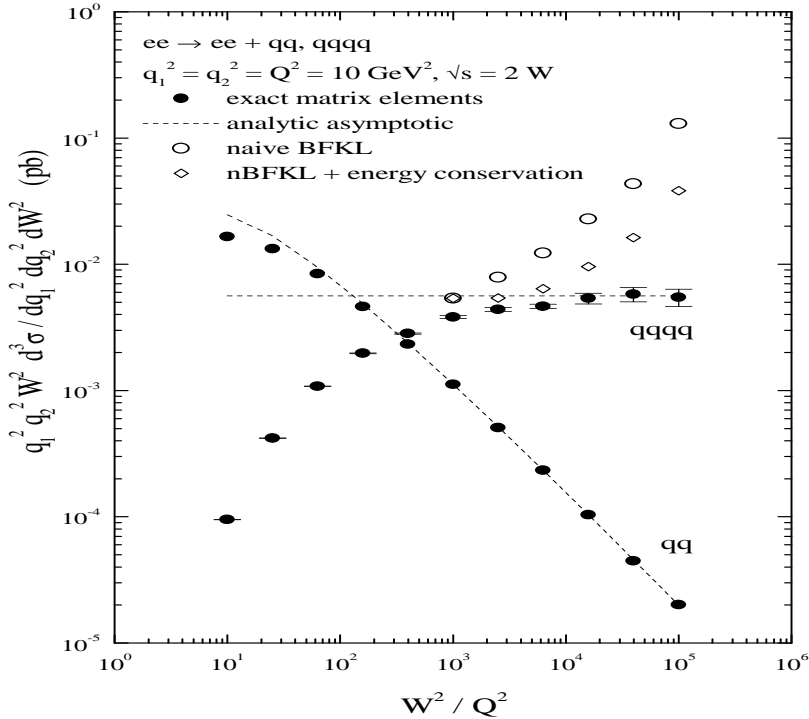,width=12cm,height=10.5cm}
\caption{Exact (closed data points) and analytic asymptotic (dashed line) 
$e^+e^- \to e^+e^-q \bar q$  and 
$e^+e^- \to e^+e^-q \bar q q \bar q$  cross sections versus 
$W^2/Q^2$ at fixed $W^2/s = 1/4$.  Also shown:  analytic BFKL without (open 
circles) and with (open diamonds) energy conservation imposed.} 
\vspace*{-.5cm}
\end{figure}

We have done this in \cite{os}, and incorporated
the running of $\alpha_s$ as well.  We applied the Monte Carlo
to dijet production at large rapidity separation at the Tevatron, 
improving the agreement between the BFKL prediction and experiment
for the azimuthal decorrelation between the jets.

\section*{Virtual $\gamma\gamma$ Scattering at $e^+e^-$ Colliders}

Returning to virtual photon scattering at linear colliders, we also consider
the leading order QCD processes from which the BFKL must be built.  It
is convenient to consider the differential cross section
$$
W^2 Q_1^2 Q_2^2 \; {d^3 \sigma \over  d W^2 d Q_1^2 d Q_2^2 } 
$$
as a function of $W^2/Q^2$.  As shown in Figure 1, the four-quark cross section 
approaches a constant as $W^2/Q^2$ gets large, but the analytic BFKL 
solution (open circles) rises.   Also shown (open diamonds) is BFKL with 
energy conservation imposed on the emitted gluons, but without the 
entire machinery of the improved BFKL Monte Carlo; the result shown can
be considered an upper limit on the improved BFKL result.  An important
point to note here is that the origin (in $W^2/Q^2$) of the BFKL curve,
viz., the point where it begins to rise from the asymptotic QCD result,
is arbitrary at leading order.  We have made what seems a reasonable 
choice (where the QCD result is approaching its asymptotic value), 
but the entire BFKL curve can in principle be shifted horizontally.

Experimentally we are more interested in the cross section for fixed 
$\sqrt{s}$; this is shown as a function of $W$ in Figure 2 for 
$\sqrt{s}=500\ {\rm GeV}$.  In this case all of the cross sections fall
off at large $W$, but the BFKL cross section lies above the others.  
The BFKL Monte Carlo calculation is in progress \cite{osprog}.

\begin{figure}
\vspace*{-1.5cm}
\epsfxsize200pt
\psfig{figure=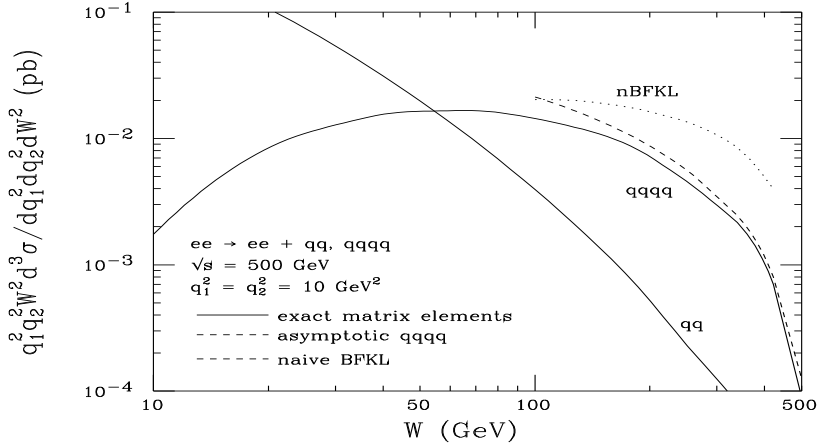,width=12cm,height=13cm}
\vspace*{-2.5cm}
\caption{Exact (solid lines) and analytic asymptotic (dashed line) 
$e^+e^- \to e^+e^-q \bar q$  and 
$e^+e^- \to e^+e^-q \bar q q \bar q$  cross sections versus 
$W^2/Q^2$ at fixed $\sqrt{s}=500\ {\rm GeV}$.  Also shown:  analytic BFKL
(dotted line).}  \label{fig:lc}
\end{figure}

\section*{Comments on LEP Results}

Finally, we note that this process has been investigated at LEP.  The 
L3 collaboration  have measured the $\gamma^*\gamma^*$ cross
section from double-tagged $e^+e^-$ events, and their
result lies between asymptotic QCD (which is flat) and analytic
BFKL (which rises); see for example\cite{LEP}.  It is likely
that the BFKL Monte Carlo prediction will lie closer to the 
data, but we note that the flat QCD prediction is the {\it asymptotic} 
result.  In fact the QCD prediction is rising in this region and has not 
reached the asymptotic limit.  Until 
the fixed-order QCD and BFKL Monte Carlo predictions are sorted out,
it is not clear exactly what we can  conclude from the data.
This work, at both LEP and  
linear collider energies, is currently in progress\cite{osprog}.

\end{document}